\documentclass[epjCONF]{svjour}
\usepackage{graphics}
\usepackage[varg]{txfonts} 
\usepackage[latin1]{inputenc}
\newcommand{\co}{\mbox{C$/$O}}		
\newcommand{\aovhe}{\mbox{$\alpha_{\mbox{\tiny ovhe}}$}}
\newcommand{\gradT}{\mbox{$\nabla_{\mathrm{T}}$}}	
\newcommand{\gradr}{\mbox{$\nabla_{\mathrm{rad}}$}}
\newcommand{\grada}{\mbox{$\nabla_{\mathrm{ad}}$}}

\newcommand{\Msun}{\mbox{M$_{\odot}$}}
\session-title{Conference Title, to be filled}
\begin{document}
\title{The AGB bump: a calibrator for core mixing}
\author{Diego Bossini\inst{1,2}\fnmsep\thanks{\email{dbossini@bison.ph.bham.ac.uk}} 
   \and Andrea Miglio\inst{1,2} 
   \and Maurizio Salaris\inst{3} 
   \and L\'eo Girardi\inst{4} 
   \and Josefina Montalb\'an \inst{5} 
   \and Alessandro Bressan\inst{6} 
   \and Paola Marigo\inst{5}
   \and Arlette Noels\inst{7}}
\institute{School of Physics and Astronomy, University of Birmingham, UK
      \and Stellar Astrophysics Centre, Department of Physics and Astronomy, Aarhus University, Aarhus, DK
      \and Astrophysics Research Institute, Liverpool John Moores University, UK
      \and INAF--Osservatorio Astronomico di Padova, Padova, IT
      \and Dipartimento di Fisica e Astronomia, Universit\`a di Padova, Padova, IT
      \and Scuola Internazione Superiore di Studi Avanzati SISSA, Trieste, IT
      \and Institut d'Astrophysique et G\'eophysique de Universit\'e de Li\`ege, Li\`ege, BE
}
\abstract{
The efficiency of convection in stars affects many aspects of their evolution and remains one of the key-open questions in stellar modelling.
In particular, the size of the mixed core in core-He-burning low-mass stars  is still uncertain and impacts the lifetime of this evolutionary phase and, e.g., the \co\ profile in white dwarfs.
One of the known observables related to the Horizontal Branch (HB) and Asymptotic Giant Branch (AGB) evolution is the AGB bump. Its luminosity depends on the position in mass of the helium-burning shell at its first ignition, that is affected by the extension of the central mixed region.
In this preliminary work we show how various assumptions on near-core mixing and on the thermal stratification in the overshooting region affect the luminosity of the AGB bump, as well as the period spacing of gravity modes in  core-He-burning  models.}

\maketitle
\section{Stellar Models}
\label{models}
We have used the code MESA (Modules for Experiments in Stellar Astrophysics, \cite{Paxton_etal11}) to study the effects of different near-core mixing prescriptions during  the helium-burning phase on the luminosity of the AGB bump, and on the period spacing of gravity modes.
At this stage, we have explored two extreme cases: no extra mixing and very large extra-mixing (see \cite{Straniero_etal03}). We computed three evolutionary sequences of models of same mass ($M=1.5$ \Msun) and chemical composition ($Z = 0.0176$, $Y = 0.266$), but changing the  near-core mixing scheme:
\begin{enumerate}
\item Bare-Schwarzschild (BS): no semiconvection, no breathing pulses, which leads to an underestimation of the convective-core size (\cite{Castellani_etal71a}, \cite{Gabriel_etal14});
\item High Overshooting (OV): step function overshooting,  using the Maeder \& Meynet 1987 scheme (\cite{Maeder&Meynet87}) with the overshooting parameter $\aovhe=1\mathrm{H}_p$ (extention in radius from the classic border) and the radiative gradient of temperarure in the overshooting region ($\gradT_\mathrm{ovhe} = \gradr$);
\item Penetrative Convection (PC) : step function overshooting with $\aovhe = 1\mathrm{H}_p$ and the adiabatic gradient of temperarure in the overshooting region ($\gradT_\mathrm{ovhe} = \grada$).
\end{enumerate}

\section{Effects on the Luminosity of the AGB Bump}
\label{sec:L_age}
In models computed with extra-mixing  (OV and PC) the luminosity of the AGB bump is higher than in the BS model  (see Figure \ref{fig1}). The total lifetime (HB plus AGB) changes significantly when this extra mixing is considered (increasing by about 40\% compared to the BS model).  However, looking at the lifespan of the single phases, what increases substantially is the duration of the core-burning phase, while that of the AGB (and the AGB bump) decreases (see \cite{Bressan_etal86}). 

\section{Effects on the Period Spacing of Core-He-Burning Models}
\label{sec:DP_Y}
\begin{figure}
    \begin{minipage}[t]{0.45\columnwidth}
        \centering
        \resizebox{\columnwidth}{!}{\includegraphics{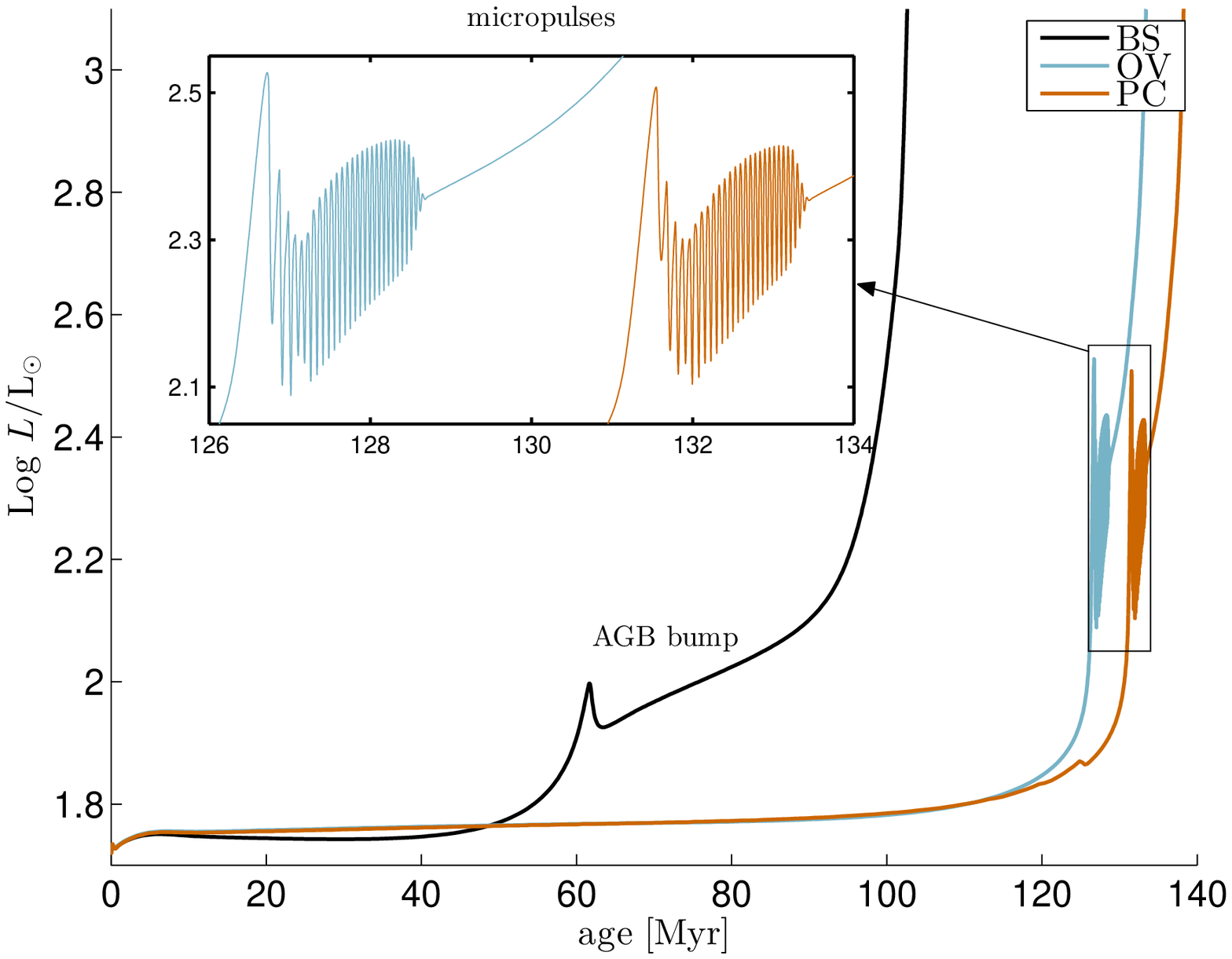}}
        \caption{Luminosity as a function of age from the start of central He-burning for the three models described in the main text. Large extra-mixing models show a complex feature during the AGB bump, known as {\it micropulses} (see \cite{Mazzitelli_DAntona86}).}
        \label{fig1}
    \end{minipage}
        \ \hspace{0.1\columnwidth}\
    \begin{minipage}[t]{0.45\columnwidth}
        \centering
        \resizebox{\columnwidth}{!}{\includegraphics{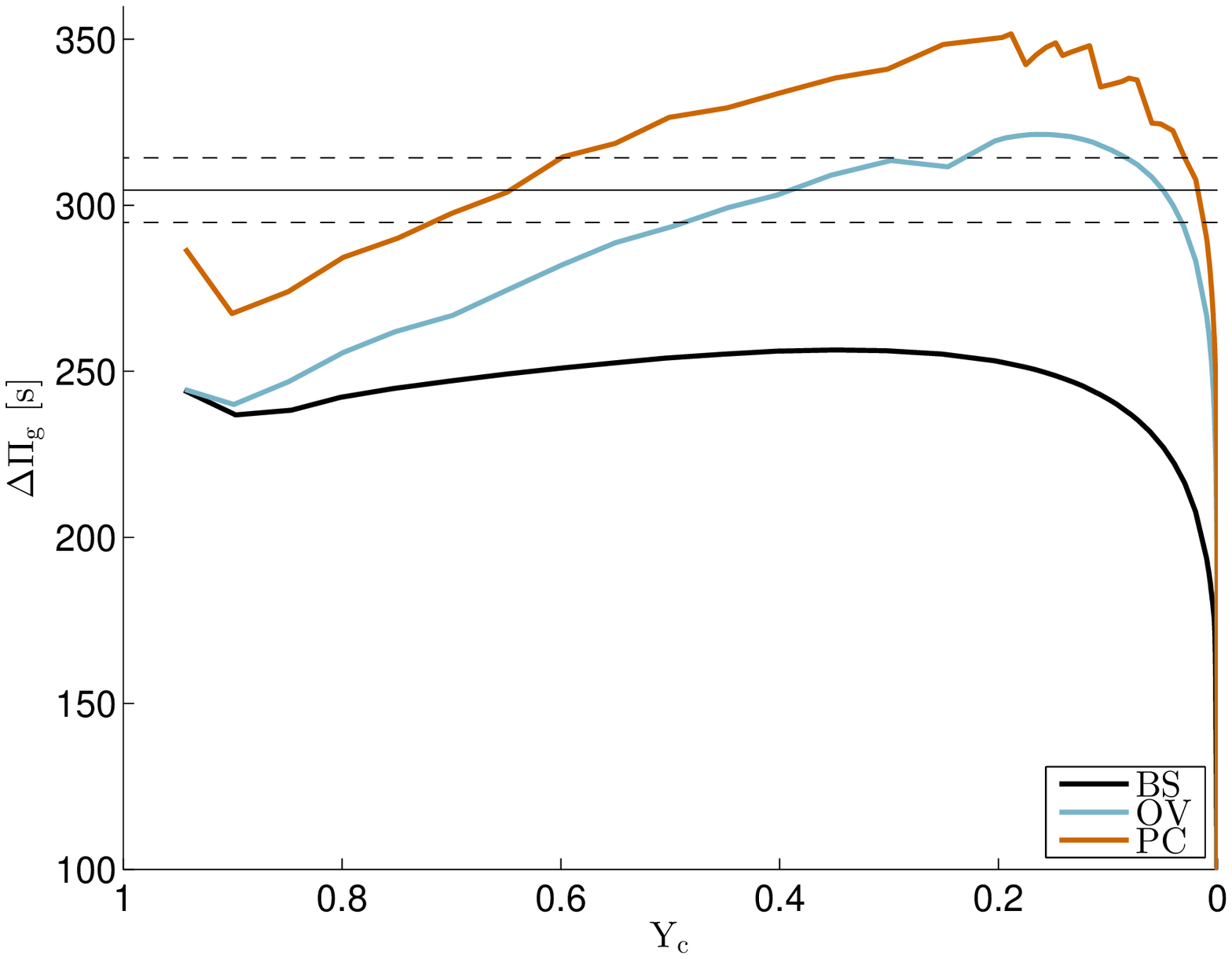}}
        \caption{Period spacing as function of central Helium mass fraction for the same models shown in Fig. \ref{fig1}. The horizontal solid line represents the average period spacing of a sample of  {\it Kepler} giants \cite{Mosser_etal12}.}
        \label{fig2}
    \end{minipage}
\end{figure}

Models (and stars, see \cite{Bedding_etal11}) in the core-helium burning phase show a larger period spacing than while on the red-giant branch. As shown in Fig. \ref{fig2}, the period spacing of He-burning models increases even more when considering extra mixing  (see also \cite{Montalban_etal13}). Moreover, we found a similar increase in period spacing when comparing PC and OV models, due to the fact that the choice of the temperature gradient in the overshooting region has a direct impact on the Brunt-V\"{a}is\"{a}l\"{a} frequency $N$:

\begin{itemize}
\item PC models have $N^2_\mathrm{ovhe} \propto \nabla_\mathrm{ad}-\nabla_\mathrm{T}=\nabla_\mathrm{ad}-\nabla_\mathrm{ad}=0$
\item OV models have $N^2_\mathrm{ovhe} \propto \nabla_\mathrm{ad}-\nabla_\mathrm{T}=\nabla_\mathrm{ad}-\nabla_\mathrm{rad}>0$.
\end{itemize}
Since the the period spacing of gravity modes is inversely proportional to $\int{\frac{N}{r}}{\rm}dr$, PC models have a larger period spacing than OV models. \\

Further details and a comparison with observational constraints (period spacing of gravity modes and  luminosity of the AGB bump) will be presented a forthcoming paper. 
\bibliographystyle{epj}
\bibliography{Bossini_D}


\end{document}